\input harvmac


\def\IZ{\relax\ifmmode\mathchoice  
{\hbox{\cmss Z\kern-.4em Z}}{\hbox{\cmss Z\kern-.4em Z}}  
{\lower.9pt\hbox{\cmsss Z\kern-.4em Z}}  
{\lower1.2pt\hbox{\cmsss Z\kern-.4em Z}}\else{\cmss Z\kern-.4em  
Z}\fi} 

\def\bZ{\bf{Z}}
\def\frac#1#2{{#1 \over #2}}
\def\dr{\rangle\rangle}
\def\dl{\langle\langle}
\def\xo{x_{0}}
\def\txo{\tilde{x}_0}

\def\talpha{\tilde\alpha}
\def\tJ{\tilde J}
\def\tX{\tilde X}

\def\bz{\bar z}
\def\bw{\bar w}
\def\pa{\phi_1}
\def\pb{\phi_2}
\def\lam{\lambda}
\def\ep{\epsilon}
\def\Jo{J_0}
\def\Djmn{D^j_{m,n}}
\def\gb{g_b}
\def\hp{\hat{p}}
\def\npb#1#2#3{Nucl.\ Phys. {\bf B#1} (19#2) #3}
\def\npbm#1#2#3{Nucl.\ Phys. {\bf B#1} (20#2) #3}
\def\plb#1#2#3{Phys.\ Lett. {\bf B#1} (19#2) #3}
\def\prl#1#2#3{Phys.\ Rev.\ Lett. {\bf #1} (19#2) #3}
\def\prd#1#2#3{Phys.\ Rev. {\bf D#1} (19#2) #3}

\def\mpla#1#2#3{Mod.\ Phys.\ Lett. {\bf A#1} (19#2) #3}
\def\cmp#1#2#3{Commun.\ Math.\ Phys. {\bf #1} (19#2) #3}

\def\jhep#1#2#3{JHEP {\bf #1} (#2) #3}
\lref\sen{A.~Sen, ``Descent Relations Among Bosonic D-branes,'' Int.\
J.\ Mod.\ Phys. {\bf A14} (1999) 4061, hep-th/9902105.}
\lref\ck{C.~G.~Callan, I.~R.~Klebanov, ``Exact c=1 Boundary Conformal
Field Theories,'' \prl{72}{94}{1968}, hep-th/9311092.}
\lref\cklm{C.~G.~Callan, I.~R.~Klebanov, A.~W.~W.~Ludwig, and
J.~M.~Maldacena, ``Exact Solution of a Boundary Conformal Field
Theory,'' \npb{422}{94}{417}, hep-th/9402113.}
\lref\poth{J.~Polchinski and L.~Thorlacius, ``Free Fermion Representation
of a Boundary Conformal Field Theory,'' \prd{50}{94}{622}, hep-th/9404008.}
\lref\gr{M.~R.~Gaberdiel and A.~Recknagel, ``Conformal Boundary States
for Free Bosons and Fermions,'' \jhep{0111}{2001}{016}, hep-th/0108238.}
\lref\jan{R.~J.~Janik, ``Exceptional Boundary States at c=1,''
\npbm{618}{01}{675}, hep-th/0109021.}
\lref\rs{A.~Recknagel and V.~Schomerus, ``Boundary Deformation Theory
and Moduli Spaces of D-Branes,'' \npb{545}{99}{233}, hep-th/9811237.}
\lref\fri{D.~Friedan, ``The Space of Conformal Boundary Conditions for
the c=1 Gaussian Model,'' unpublished (1999).}
\lref\kp{I.~R.~Klebanov and A.~M.~Polyakov, ``Interaction of Discrete
States in Two-Dimensional String Theory,'' \mpla{6}{91}{3273}, hep-th/9109032.}
\lref\car{J.~L.~Cardy, ``Boundary Conditions, Fusion Rules and the
Verlinde Formula,'' \npb{324}{89}{581}.}
\lref\cl{J.~L.~Cardy and D.~C.~Lewellen, ``Bulk and Boundary Operators
in Conformal Field Theory,'' \plb{259}{91}{274}.}
\lref\al{I.~Affleck and A.~W.~W. Ludwig,
``Universal Noninteger Ground State Degeneracy In Critical Quantum
Systems,'' \prl{67}{91}{161}.}
\lref\alu{I.~ Affleck and A.~W.~W.~Ludwig, ``  ,'' Phys.\ Rev. {\bf B48} 
(1993) 7297.}
\lref\kmm{D.~Kutasov, M.~Marino, and G.~Moore, ``Some Exact Results on
Tachyon Condensation in String Field Theory,'' \jhep{0010}{2000}{045},
hep-th/0009148.}
\lref\lew{D.~C.~Lewellen, ``Sewing Constraints for Conformal Field
Theories on Surfaces with Boundaries,'' \npb{372}{92}{654}.}
\lref\gp{E.~G.~Gimon and J.~Polchinski, ``Consistency Conditions for
Orientifolds and D-manifolds,'' \prd{54}{96}{1667}, hep-th/9601038.}
\lref\dm{M.~R.~Douglas and G.~Moore, ``D-branes, Quivers, and ALE
Instantons,'' hep-th/9603167.}
\lref\dou{M.~R.~Douglas, ``Two Lectures on D-Geometry and Noncommutative
Geometry,'' hep-th/9901146.}
\lref\mamo{E.~J.~Martinec and G.~Moore, ``Noncommutative Solitons on
Orbifold,''~ hep-th/0101199.}
\lref\grw{M.~R.~Gaberdiel, A.~Recknagel, and G.~M.~T.~Watts, ``The
Conformal Boundary States for SU(2) at Level 1,'' \npbm{626}{02}{344},
hep-th/0108102.}
\lref\dg{D.-E.~Diaconescu and J.~Gomis, ``Fractional Branes and Boundary
States in Orbifold Theories,'' \jhep{0010}{2000}{001}, hep-th/9906242.}
\lref\gs{M.~R.~Gaberdiel and B.~Stefanski~Jr., ``Dirichlet Branes on
Orbifolds,'' \npbm{578}{00}{58}, hep-th/9910109.}
\lref\hkms{J.~A.~Harvey, S.~Kachru, G.~Moore, and E.~Silverstein,
``Tension is Dimension,'' \jhep{0003}{2000}{001}, hep-th/9909072.}
\lref\sw{N.~Seiberg and E.~Witten, ``String Theory and Noncommutative
Geometry,'' JHEP {\bf 9909} (1999) 032, hep-th/9908142.}
\lref\conn{A.~Connes, {\it Noncommutative Geometry,} Academic Press,
(1994);~A.~Connes, "Noncommutative Geometry 2000," math.QA/0011193.}
\lref\nsv{K.~S.~Narain, M.~H.~Sarmadi, and C.~Vafa, ``Asymmetric
Orbifolds,'' \npb{288}{87}{551}.}
\lref\bcr{M.~Billo, B.~Craps, and F.~Roose, ``Orbifold Boundary States
from Cardy's Condition,'' \jhep{0101}{2001}{038}, hep-th/0011060.}
\lref\capdap{A.~Cappelli and G.~D'Appollonio, ``Boundary States of $c=1$
and $3/2$ Rational Conformal Field Theories,'' \jhep{0202}{2002}{039},
hep-th/0201173.}
\lref\msak{M.~Sakamoto, ``A Physical Interpretation of Cocycle Factors in
Vertex Operator Representations,'' \plb{231}{89}{258}.}
\lref\ghmr{D.~J.~Gross, J.~A.~Harvey, E.~J.~Martinec and R.~Rohm,
``Heterotic String Theory (1). The Free Heterotic String,'' \npb
{256}{85}{253}.}
\lref\godoliv{P.~Goddard and D.~Olive, ``Kac-Moody and Virasoro Algebras in
Relation to Quantum Physics,''  Int.\ J.\ Mod.\ Phys. {\bf A1} (1986)
303;~ P.~Goddard, W.~Nahm, D.~Olive, and A.~Schwimmer, ``Vertex Operators
for Non-Simply-Laced Algebras,'' \cmp{107}{86}{179}.}
\lref\pol{J.~Polchinski, {\it String Theory, volume 1: An Introduction to
the Bosonic String,} Cambridge University Press, (1998).}
\Title{\vbox{\baselineskip12pt
\hbox{hep-th/0201254}
\hbox{EFI-02-58}
\vskip-.5in}}
{\vbox{\centerline{A Note on $c=1$ Virasoro Boundary States and}
\centerline{Asymmetric Shift Orbifolds}}}
\medskip\medskip\bigskip
\centerline{Li-Sheng Tseng}
\bigskip\medskip
\centerline{\it Enrico Fermi Institute and Department of Physics}
\centerline{\it  University of Chicago,} 
\centerline{\it  5640 S. Ellis Ave., 
Chicago, IL 60637, USA}
\centerline{\tt lstseng@theory.uchicago.edu}
\medskip
\baselineskip18pt
\medskip\bigskip\medskip\bigskip\medskip
\baselineskip16pt
\noindent
We comment on the conformal boundary states of the $c=1$ free boson theory on a
circle which do not preserve the U(1) symmetry.  We construct these Virasoro
boundary states at a generic radius by a simple asymmetric shift orbifold
acting on the fundamental boundary states at the self-dual
radius.  We further calculate the boundary entropy and find that the
Virasoro boundary states at irrational radius have infinite boundary
entropy.  The corresponding open string description of the asymmetric
orbifold is given using the quotient algebra construction.  Moreover, 
we find that the quotient algebra associated with a non-fundamental
boundary state contains the noncommutative Weyl algebra.

\Date{January, 2002}
\newsec{Introduction}

Recently, there has been renewed interest in the conformal boundary states of
theories with central charge c=1  \refs{\fri, \gr, \jan, \capdap}.  For the c=1
free boson theory taking values on a circle with arbitrary radius $R$,
there are two well-known classes of boundary states corresponding to
Dirichlet and Neumann boundary conditions.  Considering only one spatial
direction in the string theory framework, these conformal boundary conditions
correspond to $D0$-branes and $D1$-branes, respectively.  Each brane has
a classical modulus - for the $D0$-brane, the value
of its location, and for the $D1$-brane, the value of the Wilson line.

It has also been known that there exist
additional conformal boundary states for $R=M\ {\rm or}\
1/M$, integer multiples (or one over integer multiples) of the
self-dual radius, $R_{sd}=\sqrt{\alpha'}= 1$  \refs{\ck, \cklm, \poth,
\rs}.  These conformal boundary conditions can not be simply
expressed using only the left and right U(1) currents.  They correspond
in string theory to the addition of a marginal tachyon potential on the
boundary of the worldsheet action.  The marginal tachyon potential breaks the
extended U(1) current algebra but still preserves the Virasoro algebra.
Recent works have now given explicit constructions of non-U(1)
conformal boundary states for any rational \gr\ and irrational radius
\refs{\gr, \jan}.

Central to the construction of all non-U(1) boundary states is the
presence of the non-U(1) discrete state primaries in the c=1 CFT (see \kp\ for
details and references).  The U(1) representation with the
highest weight primary $e^{ipX(z)}$ is reducible under the Virasoro
algebra when the conformal dimension $h=\frac{p^2}{4}=j^2$ where $j\in
\frac{\bZ}{2}$ (and similarly for the anti-holomorphic sector).  Thus,
for integer values of $p$, there are U(1) descendants that are Virasoro
primaries.  Given that the left and right momenta at $R=1$ have values
$(p_L,p_R)=(n+m,n-m)$ for $m,n\in \bZ$, 
all discrete primaries are found in the $R=1$ self-dual theory
and are organized into multiplets of the SU(2)$\times$SU(2) enhanced
symmetry.  These discrete primaries and their descendants can then be
grouped together to form Virasoro Ishibashi states.  The non-U(1)
boundary states are then a linear combination of Virasoro Ishibashi
states with coefficients chosen to satisfy Cardy's condition \car .  We
will call all $c=1$ conformal boundary states that can only be
constructed using Virasoro Ishibashi states Virasoro boundary states.

Due to the discreteness of the momenta $(p_L,p_R)=(n/R +
mR, n/R-mR)$, theories at different radii have different sets of Virasoro
Ishibashi states.  However, it is important to emphasize that all
possible Virasoro Ishibashi states are present in the $R=1$ theory.  This
suggests that there may be a connection between the Virasoro boundary states
constructed at $R\neq 1$ to those at $R=1$.  Below we will show that the
connection at a generic radius $r$ is that of an asymmetric shift orbifold.
With the shift acting differently on the left- and right-movers, the
closed string theory at $R=1$ can be orbifolded to the closed string
theory at any other radius.  Thus, all Virasoro boundary states at
$R\neq 1$ can be obtained by the orbifold boundary state construction.
Effectively, the orbifold action on the $R=1$ boundary states projects
out Virasoro Ishibashi states in the $R=1$ theory that are not
present in the $R=r$ theory. 

In section 2, we introduce our notation by
reviewing the construction of $R=1$ Virasoro boundary states.  We
present in section 3 the asymmetric shift orbifold and the boundary
state orbifold construction connecting Virasoro boundary states at
different radii.  We then proceed to analyze the orbifold boundary
states for their algebraic and geometrical significance. 
In section 4, we describe the boundary state orbifold operation in the
open string framework.  Here, the analogous prescription is given by the
quotient algebra construction.  For the $D1$- and $D0$-brane boundary
conditions at a generic radius, we note the presence of the noncommutative
Weyl algebra in the quotient algebra.  In section 5, we conclude with a
discussion generalizing the orbifold construction of branes and
propose a condition for establishing a connection between boundary
field theories situated within different bulk theories.

\newsec{R=1 Virasoro boundary states}

Before generating the Virasoro boundary states for $R\neq 1$, we
introduce our notation by writing down the U(1) and Virasoro boundary
states at $R=1$ (see \refs{\rs , \grw} for some of the details).  As
mentioned above, the Virasoro primaries are organized into
SU(2)$\times$SU(2) multiplets with the SU(2) currents given by
\eqn\sugen{J^{x}(z)=\cos 2 X_{L}(z)~, \qquad J^{y}(z)=\sin 2 X_{L}(z)~, \qquad
 J^{z}(z)=i\partial X_{L}(z) ~,}
in the holomorphic sector (and similarly in the anti-holomorphic sector).
For constructing boundary states, we are mainly
interested in the spin zero primaries.  These can be generated from the
highest weight states $e^{i\,2j(X_L(z)+X_R(\bz))}$ where $X(z,\bz)\equiv
X_L(z)+X_R(\bz)$ and $j=0,\frac{1}{2},1,\ldots$ .  Acting with
lowering operators $J_0^{-}(z)$ and $\tJ_0^{-}(\bz)$, the spin zero
primaries are explicitly
\eqn\virprim{\eqalign{\phi^j_{m,n}(z,\bz)=
\frac{1}{(2j)!}&\sqrt{\frac{(j+m)!(j+n)!}{(j-m)!(j-n)!}}
\; \left [\oint_{\cal C} {dw\over 2\pi i} \;e^{-i\, 2X_L(w)}
e^{i\frac{\pi}{2}\hp_R}\right ]^{j-m}\cr &\times \left [-\oint_{\cal C} {d\bw\over 2\pi i} \;e^{-i\,
2X_R(\bw)} e^{-i\frac{\pi}{2}\hp_L}
\right ]^{j-n} e^{i\,
2j(X_L(z)+X_R(\bz))}e^{i\frac{\pi}{2}j(\hp_L-\hp_R)} ~, }}
where we have included the appropriate normalization factor and also cocycle
factors which have dependence on the left and right momentum
operators, $\hp_L=2J^z_0$ and $\hp_R=2\tJ^z_0$, respectively.  The
cocycle factor $c_k(\hp)=e^{i\frac{\pi}{4}(k_R\hp_L-k_L\hp_R)}$ is attached
to the right of each vertex operator with zero mode $e^{ik_LX_L(z)+ik_RX_R(\bz)}$.  The operator $c_k(\hp)$ satisfies 
\eqn\cocy{c_k(\hp+k')c_{k'}(\hp)=(-1)^{\frac{1}{2}(k_Lk'_L-k_Rk'_R)}c_{k'}(\hp+k)c_{k}(\hp)=\ep(k,k')c_{k+k'}(\hp)~,}
with the two-cocycle $\ep(k,k')=e^{i\frac{\pi}{4}(k_Rk'_L-k_Lk'_R)}\,$.
The cocycles are needed to insure that the vertex operators commute
amongst themselves \pol .  We point out that the expression for
the cocycle factor is not unique.  Our choice of $c_k(\hp)$ preserves
the commutation relations of the SU(2)$\times$SU(2) current
algebra.\foot{For additional details on two-cocycles, see \refs{\ghmr,
\godoliv, \msak}.}  Besides the cocycle, the primaries
$\phi^j_{m,n}(z,\bz)$ in \virprim\ have $h={\tilde h}=j^2$ and are
polynomials in derivatives of $X$ multiplied by the zero mode,
$e^{i\,2mX_L+i\,2nX_R}$.   We will label the Virasoro Ishibashi states
generated by the primaries $\phi^j_{m,n}(z,\bz)$ by $|j;m,n\dr$.   
  
The Dirichlet $D0$-brane boundary state is typically written as a sum
over U(1) Ishibashi states with $p_{L}=p_{R}$.  In particular, a
$D0$-brane located at $\xo=0$ for $R=1$ can be expressed as
\eqn\Dxo{\eqalign{|D0\dr_{\xo=0}&=\frac{1}{2^{\frac{1}{4}}}
\sum_{n=-\infty}^{\infty}
\exp\left (\sum_{m=1}^{\infty}\frac{1}{m}\alpha_{-m}\talpha_{-m}\right )
|p_{L}\!\!=\!\!n,p_{R}\!\!=\!\!n\rangle\cr & = \frac{1}{2^{\frac{1}{4}}}
\sum_{j,m} (-1)^{j-m} |j;m,m\dr~,}}
where in the second line we have written the boundary
state as a sum over Virasoro Ishibashi states.\foot{Note that the
presence of cocycle changes the definition of $\phi^j_{m,n}(z,\bz)$ up
to a phase.  This becomes important in matching the U(1) and Virasoro
primaries in \Dxo\ and checking the $D0$-brane boundary
conditions.}$^,$\foot{Some authors, as for example in \cklm , normalize
$|j;m,n\dr$ so that the factor $(-1)^{j-m}$ is absent in the second line
of \Dxo.  Our definition of $|j;m,n\dr$ allows us to use standard
formulas for operators in SU(2) representations.}  Taking the string
worldsheet to be the upper half-plane, the $D0$-brane at $\xo=0$ implies
the following boundary conditions for the SU(2) currents at the boundary: 
\eqn\suc{ J^{x}(z)=\tJ^{x}(\bz)|_{z=\bz}~,
\qquad J^{y}(z)=-\tJ^{y}(\bz)|_{z=\bz}~, \qquad
J^{z}(z)=-\tJ^{z}(\bz)|_{z=\bz}~.} 
These conditions arise
from the general Dirichlet boundary condition $X_{L}=-X_{R}+\xo$ at
$z\!=\!\bz\,$.  As operators acting on $|D0\dr_{\xo=0}$, the boundary
conditions \suc\  can be expressed for all integer $n$ as 
\eqn\Dxb{\left (e^{i\Jo^{x}\pi}J^{a}_n e^{-i\Jo^x\pi} + \tJ^a_{-n}\right
)\ |D0\dr_{\xo=0}=0~.}

All fundamental boundary states at the self-dual radius, including
$|D0\dr_{\xo=0}\,$, contain three truly marginal boundary fields.  These
fields can be used to deform the theory and map out the space of
boundary conditions at $R=1$.  Under the deformation, 
\eqn\Dxg{|D0\dr_{\xo=0} \rightarrow |g\dr= e^{i\vec{\Jo}\cdot\hat{n}\,\phi}\
|D0\dr_{\xo=0} = \frac{1}{2^{\frac{1}{4}}}
\sum_{j,m,n}(-1)^{j-n}\, D^j_{m,n}(g)|j;m,n\dr~.}
The moduli space of fundamental boundary states is labelled by $g
\in SU(2)$ and $D^j_{m,n}(g)$ in \Dxg\ is the 
matrix element of $g$ in the $j^{\rm th}$ representation.\foot{In \grw
, it was found that the moduli space at $R=1$ can be at most SL(2,{\bf
C}).  We will not consider these possible additional fundamental
boundary states here.} We will conveniently parameterize $g$ by 
\eqn\geq{g(\lam,\pa,\pb) = \left( \matrix{\cos\lam~ e^{i\pa} & -i
\sin\lam~ e^{-i\pb} \cr -i \sin\lam~ e^{i\pb} & \cos\lam~ e^{-i\pa} }
\right) ~,}
where $\pa$ and $\pb$ are periodic in $2\pi$ and $0\leq \lam \leq
\frac{\pi}{2}\,$.

The fundamental boundary state $|g\dr$ satisfies 
\eqn\dgb{{\rm Ad}(g\cdot i) J^{a}_n + \tJ^a_{-n}\ |g\dr =0  \qquad {\rm
with} \qquad i=\left(\matrix{0&i\cr i&0} \right)~.}
By working out the explicit
boundary conditions as in \suc\ for $|D0\dr_{\xo=0}\,$, one finds that
at $\lam=0$, the boundary state satisfies the Dirichlet condition,
$J^{z}(z)=-\tJ^{z}(\bz)|_{z=\bz}$, with the $D0$-brane situated at
$\xo=R \pa = \pa$.  And at $\lam={\pi \over 2}$, the boundary state
satisfies the Neumann condition, $J^{z}(z)=\tJ^{z}(\bz)|_{z=\bz}$, with
the value of the constant gauge potential on the $D1$-brane (in units of
length) given by $\txo=\frac{\alpha'}{R} \pb = \pb$.  Here, the Neumann
boundary condition with a constant gauge potential corresponds to the
worldsheet boundary condition
$X_{L}=X_{R}+\txo$ at $z\!=\!\bz\,$. In short, we have
\eqn\dexpa{|D0\dr_{\xo}=|g(\lam=0,\pa=\xo)\dr \quad {\rm and} \quad
|D1\dr_{\txo}=|g(\lam=\frac{\pi}{2},\pb=\txo)\dr ~, } where we have
noted the independence of $g$ on the parameters $\pb$ and $\pa$
when $\lam=0$ and $\lam=\frac{\pi}{2}$, respectively.  The parameter
$\lam$ is physically the renormalized value of the strength of the
marginal tachyon potential \cklm\ .  And finally, the Virasoro boundary
states at $R=1$ are the boundary states $|g(\lam\neq 0\,{\rm
or}\,\frac{\pi}{2},\pa,\pb)\dr$. 

\newsec{Boundary State Orbifold From $R=1$ to $R=r$}

Gaberdiel and Recknagel in \gr\ constructed the Virasoro
boundary states at $R=\frac{M}{N}$ starting with the Virasoro boundary
states at $R=1$ and then projecting out those Virasoro
Ishibashi states not present at the rational radius.  We will motivate
their projection physically as arising from the construction of boundary
states in an orbifold theory.

\subsec{Closed String Orbifold to $R=r$} 

Before proceeding to orbifolding the boundary states, we first identify
the closed string orbifold connecting the theory at $R=1$ to that at a
generic radius $R=r$.  The closed string theory at any rational radius can be
obtained from that at $R=1$ by a combination of two symmetric ${\bZ}_N$ shift
orbifolds and a T-duality.  Explicitly, starting at $R=1$, we apply a ${\bZ}_M$
shift orbifold, $X(z,\bz)\equiv X_L(z)+X_R(\bz) \to X+2\pi/M$ to reduce
the radius to $R={1\over M}$. After a T-duality, we orbifold again by
${\bZ}_N$ to obtain the closed string theory at radius
$R=\frac{M}{N}$.  However, we would like to consider the procedure as a
single orbifold.  The only obstacle is the presence of a T-duality and
it can be overcome by considering orbifolding the dual field
$\tX(z,\bz)\equiv X_L(z)-X_R(\bz)$.  Take for example the ``orbifold''
to the $R=2$ theory.  The desired orbifold must project out half of the winding
states, those with vertex operator of the form $e^{i(2m+1)(X_L-X_R)}$, and
generate in the twisted sectors new momentum modes,
$(p_L,p_R)=(n+\frac{1}{2},n+\frac{1}{2})$.  By recalling the mode
expansion at radius $R$ for 
$\tX(\tau,\sigma)=X_L(z)-X_R(\bz)=\frac{x_L}{2}-\frac{x_R}{2}+\frac{n}{R}\sigma
+\, oscillator\; modes\;$, it becomes evident that the orbifold is
constructed by a shift of the dual coordinate $\tX\to\tX+2\pi/r$ for
$r=2$.  In particular, the orbifold can be written explicitly as $X_L\to
X_L + \pi/2$ and $X_R\to X_R - \pi/2$; thus, it is a simple example of
an asymmetric orbifold \nsv .

Now generalizing the orbifold to arbitrary rational radius $r$, we find
that the required asymmetric orbifold group action $\Gamma$ is generated by the
following two elements: $X\to X+2\pi r$ and $\tX\to \tX+
\frac{2\pi}{r}$.  Physically, the generators reset the periodicities of the
coordinate and the dual coordinate to that required at the $R=r$ theory.
(Recall that at $R=1$, both $X$ and $\tX$ have periodicities of $2\pi$.)  We
will label a group element $h\in\Gamma$ by two indices $(m',n')$ with 
action on the fields given by 
\eqn\geh{h_{m'\!,n'}: \quad X\to X+2\pi r m'\;,\quad
\tX\to\tX+2\pi\frac{n'}{r}~.}
The identity element is $h_{0,0}$ and the order of the group action is given
by $|\Gamma|\equiv {(m_{max}'+1)(n_{max}'+1)}$.

The orbifold group action
$\Gamma={\bZ}_{(m_{max}'+1)}\times{\bZ}_{(n_{max}'+1)}$ with elements
given in \geh\ can in
fact be applied to any irrational radius $r$.  The only difference is
that $|\Gamma|$ is now infinite with both $m_{max}'$ and $n_{max}'$
taken to infinity.  The orbifold theory is the closed string theory at
the irrational radius $r$.  Here, only states with zero left and right
momenta in the $R=1$ theory are invariant under the orbifold projection.
The momentum mode $(p_L,p_R)=(n'/r+m'r,n'/r-m'r)$ and its conformal
descendants are generated by the twisted sector associated with
$h_{m'\!,n'}$.  Thus, setting $m_{max}'=n_{max}'=\infty$, the orbifold
partition function will contain all the left and right momentum
combinations of the $R=r$ theory.  All together, we have obtained an
asymmetric shift orbifold that allows us to orbifold the $R=1$ closed
string theory to that at any arbitrary radius.  In the following, for
ease of description, we will call the theory at $R=1$ the ``covering
space'' theory and that at $R=r$, the orbifold space theory.

\subsec{Constructing Virasoro Orbifold Boundary State} 

In constructing an orbifold boundary state, one first identifies the 
physical parameter in which the orbifold group action $\Gamma$ acts.  Then the
``untwisted'' or bulk orbifold boundary states can be constructed from
the covering space by summing over all images of the boundary state
under the orbifold group action and dividing the sum by the
normalization factor, $\sqrt{|\Gamma|}\,$.  This prescription ensures that
the resulting boundary state is $\Gamma$ invariant as the non-invariant
closed string states are effectively projected out.  For
the simple example of orbifolding an $S^{1}$ theory by the ${\bZ}_2$
shift action $X\to X+2\pi R/2$, the $D0$-brane bulk orbifold boundary
state is given by $\frac{1}{\sqrt 2}\left (|D0\dr_{\xo}+|D0\dr_{\xo+2\pi
R/2} \right )$.  Here, the parameter which is acted upon is the location
$\xo$ of the $D0$-brane.

For orbifolding to the $R=r$ theory, we note that the boundary states at
$R=1$ are parameterized by $(\lam, \pa, \pb)$.  In changing the radius $R$,
the location of the $D0$-brane, $\pa$, and the value of
the constant gauge potential, $\pb$, must now satisfy new periodicities.  The
orbifold group element $h_{m'\!,n'}$ in \geh\ acts on the parameters as
follows: $\pa \to \pa + 2\pi r m' ~{\rm and}~ \pb \to \pb +
2\pi\frac{n'}{r}\,$.  $\lam$ being the coupling strength of the tachyon
potential is invariant under
the orbifold action. The Virasoro orbifold boundary state can then be expressed
as
\eqn\pbs{\eqalign{|g(\lam,\pa,\pb)\dr'&=\frac{1}{\sqrt{|\Gamma|}}
\sum_{m',n'}|g(\lam,\pa+2\pi r m', \pb+2\pi\frac{n'}{r})\dr\cr
&=\frac{1}{2^{\frac{1}{4}}\sqrt{|\Gamma|}}\sum_{m',n'}\sum_{j,m,n}{(-1)^{j-n}}{D^{j}_{m,n}[g(\lam,\pa+2\pi
r m', \pb+2\pi\frac{n'}{r})]}|j;m,n\dr ~,}}
where the sums over $m'$ and $n'$ is from zero to $m'_{max}$ and
$n'_{max}$, respectively,  with 
${(m_{max}'+1)(n_{max}'+1)}=|\Gamma|$.  Note that $|\Gamma|$ is
determined by $r$
and is independent of $\lam\,$.  For $r=M/N$, \pbs\
is equivalent to the projection prescription of the Virasoro boundary
states given in \gr .  In a different framework, Friedan \fri\ has also
proposed a projection mechanism in describing the space of $c=1$ boundary
conditions.  

We point out that the bulk orbifold boundary states at $R=r$ in \pbs\ are
expressed only in terms of closed string states already present in the $R=1$
theory.  There are also fundamental orbifold boundary states that
utilize twisted sector closed strings.  These are called fractional
boundary states (see for example \refs{\dg, \bcr} and
references therein).  For the asymmetric orbifold \geh , they correspond to the
the well-known single $D0$- or single $D1$-brane boundary state in the
orbifold space and will not be much discussed below.\foot{The
expression for a 
single $D0$-brane boundary state at arbitrary radius can be found in
subsection 3.4.  That for a single $D1$-brane boundary state can be
obtained by T-dualizing the single $D0$-brane expression.}  In the next subsection, we will make some general remarks
concerning the orbifold construction independent of the radius $r$.  We
will then proceed to provide some geometrical intuition for the orbifold
boundary states on the orbifold space.

\subsec{Consistency of the Orbifold Boundary States and Boundary Entropy}

The content of \pbs\ is that it expresses Virasoro boundary state of the $R=r$
theory in terms of a linear combination of Virasoro boundary states of
the $R=1$ theory.  Therefore, one should check that the resulting
boundary state consists only of the Virasoro Ishibashi states present at
the $R=r$ theory.  This is easily demonstrated by using the relations
\eqn\dgpa{\Djmn(\lam,\pa+\varphi,\pb)=e^{i\varphi(m+n)}\Djmn(\lam,\pa,\pb)~,}
\eqn\dgpb{\Djmn(\lam,\pa,\pb+\varphi)=e^{-i\varphi(m-n)}\Djmn(\lam,\pa,\pb)~.}
The sum over $m'$ and $n'$ in \pbs\ together with \dgpa\ and \dgpb\
imply that the Virasoro Ishibashi state $|j;m,n\dr$ will have a
non-zero contribution if and only if 
\eqn\vicond{(m+n)r=Z_1\qquad{\rm and}\qquad (m-n)/r=Z_2~,} 
where $Z_1\, ,\, Z_2 \in \bZ$.  Now recall that $2m$ and $2n$ are
respectively the momentum zero modes $p_L$ and $p_R$ in the discrete
state primary $\phi^j_{m,n}\,$.  Substituting
$(m,n)=(\frac{p_L}{2},\frac{p_R}{2})$ in \vicond , we arrive at the
condition that
$(p_L,p_R)=(Z_1/r\, +\, Z_2 r, Z_1/r\, -\, Z_2 r)$, precisely the
expression for the momentum modes at radius $r$.  Therefore, the
summation in \pbs\ intrinsically projects out closed string states not in
the $R=r$ theory.

The orbifold boundary states must also satisfy Cardy's condition \car
.  This is the requirement that the tree level closed strings exchange
between any two boundary states, under modular transformation, can be
expressed as the annulus partition sum of open strings in terms of
integer combination of conformal characters.  Given that the ${R=1}$
boundary states satisfy Cardy's condition, the orbifold boundary states
as a set also satisfy Cardy's condition.  Without the normalization factor
on the RHS of \pbs , Cardy's condition is trivially satisfied since an
orbifold boundary state is then just a summation of $R=1$ boundary
states.  The normalization factor just simply reduces the redundancy in
the open string conformal characters due to the summation.  More
explicitly, the closed string tree amplitude between two boundary states
$|g\dr\,,\,|g'\dr$ at $R=1$ is expressed in terms of the open string
partition sum as \gr\ 
\eqn\overl{\dl g(\lam',\pa',\pb')|q^{\frac{1}{2}(L_0+{\tilde
L}_0-\frac{c}{12}})|g(\lam,\pa,\pa)\dr=\sum_{n\in \bZ}\frac{{\tilde
q}^{\,(n+\frac{\alpha}{2\pi})^2}}{\eta(\tilde{q})} ~,}
where $\eta(\tilde{q})$ is the Dedekind $\eta$-function and $\alpha$ is
given by 
\eqn\overla{\cos \alpha\, =\, \cos\lam \cos\lam' \cos(\pa-\pa') + \sin
\lam \sin\lam' \cos(\pb-\pb')~.} 
We see that the open string conformal character for a given $\lam$ and
$\lam'$ is only dependent on the differences, $\pa-\pa'$ and
$\pb-\pb'$.  This leads to an overall multiplicity in the conformal
characters for orbifold boundary states which is divided out by the
inclusion of the normalization factor.  One still has to check that
Cardy's condition is satisfied when the fundamental fractional
boundary states containing twisted sector closed string states are also
considered.  Indeed, this is the case as can be explicitly checked by
noting that the closed string twisted sector states do not couple to the
Virasoro boundary states and by using equations \pbs,\dgpa, and \dgpb.  

It is also worthy to note that the boundary entropy $\gb$ \al , is identical
for all Virasoro boundary states at a given radius.  This is as
expected since these boundary states are connected by turning
on truly marginal boundary fields, and therefore, the boundary entropy
remains constant \refs{\al , \kmm}  .  The boundary entropy
corresponds to the coefficient of the vacuum state
$|p_L\!\!=\!0,p_R\!\!=\!0\!\!>$. 
Since the matrix element $D^0_{0,0}(g)=1$ for any $g$, we have that
\eqn\pbg{\gb=\frac{1}{2^{\frac{1}{4}}\sqrt{|\Gamma|}}\sum_{m',n'} 1
=\frac{\sqrt{|\Gamma|}}{2^{\frac{1}{4}}}~.}
Note that for rational $r=M/N$, $\gb=\frac{\sqrt{MN}}{2^{\frac{1}{4}}}$.
In contrast, for irrational $r$, the boundary entropy for all Virasoro
boundary states is infinite.  This results from the orbifold action
$\Gamma$ having infinite order at irrational $r$.  We will give a
geometrical interpretation of this infinity in the next subsection.

\subsec{Geometry of Orbifold Boundary States}

In this section, we develop some intuition on the geometry of the 
asymmetric orbifold boundary states \pbs\ in the orbifold space.
Although it is believed that the
Virasoro branes ($\,$i.e. $|g\dr$ with $0<\lam<\pi /2\,$) are
fundamental, we do not have a good geometrical picture describing a
$D1$-brane with a marginal tachyon potential turned on.  For clarity of
description, we will mostly consider the geometry of the orbifold $D0$-brane
boundary state ($\lam=0$) for arbitrary radius $r$.  The
description of the orbifold $D1$-brane state ($\lam=\frac{\pi}{2}$) is
identical to that of the $D0$-brane after a T-duality transformation.
In focusing on $D0$-branes, the Virasoro Ishibashi
states will not play a role in the description.  Therefore, our
intuition can be applied to our orbifold with the covering space
taken to have radius $R\neq 1$.  We will develop our understanding by analyzing
sequentially four generic cases: (a) $r=1/N\,$; (b) $r=N\,$; (c)
$r=M/N\,$; (d) $r={\rm irrational}\,$.

\medskip
\noindent{\it a.~ $r=1/N$ case}

This is the simplest case with the orbifold being symmetric.  The
orbifold $D0$-brane boundary state is a single $D0$-brane.  \pbs\
expresses it as a sum of its $N$ images in the $R=1$ covering space.  A
$D0$-brane has boundary entropy
$\gb=\frac{1}{2^{1\over 4}{\sqrt R}}$.
Thus, for $R=1/N$,
$\gb=\frac{\sqrt N}{2^{\frac{1}{4}}}$ which is exactly that
given in \pbg\ for $|\Gamma|=N$.  The orbifold of the $D1$-brane,
$\lam=\frac{\pi}{2}\,$, is more interesting and is described by that of
the $D0$-brane for $r=N$.

\medskip
\noindent{\it b.~ $r=N$ case}

Here, the parameter space of $\xo$ has increased in size by
$N$-fold such that $\xo \sim \xo + 2\pi N$.  The asymmetric orbifold
effectively creates copies of the original parameter space instead of
identifying the covering space.  In doing so, the orbifold
$D0$-brane boundary state should describe a configuration of $N$
evenly-spaced $D0$-branes in the $R=N$ theory. 
Indeed, this is the case as can be explicitly checked using \pbs\ and
the general boundary state expression for a single $D0$-brane,
\eqn\Dx{\eqalign{|D0\dr_{\xo}&=\frac{1}{2^{\frac{1}{4}}\sqrt R}
\sum_{n\in \bZ}e^{i\,\frac{n}{R}\xo}
e^{\sum_{m=1}^{\infty}\frac{1}{m}\alpha_{-m}\talpha_{-m}}
|p_{L}\!\!=\!\!\frac{n}{R},p_{R}\!\!=\!\!\frac{n}{R}\rangle \cr
&\equiv\frac{1}{2^{\frac{1}{4}}{\sqrt R}} \sum_{n\in
\bZ}e^{i\,\frac{n}{R}\xo}|\frac{n}{R},\frac{n}{R}\dr_D ~,}}
where in the second line we have written it in terms of U(1) Ishibashi states.
Again, the boundary entropy $\gb=N\frac{1}{2^{1\over 4}{\sqrt N}}$ is
precisely that for $N$ $D0$-branes at $R=N$.

As an aside, we use this example to raise a subtle issue
concerning the orbifold boundary state construction.  The expression for
the $D0$-brane orbifold boundary state as given in the LHS of \pbs\ is
an explicit sum of $N$ evenly-spaced $D0$-brane boundary states on the
orbifold space at $R=N$.  However, to denote a superposition of
fundamental boundary states, one should technically utilize the
Chan-Paton (CP) index instead of explicitly
summing over boundary states.  A boundary state contains
information on the couplings of closed string fields to
the open string identity fields \refs{\cl, \rs, \grw}.\foot{A field with
zero conformal dimension is by definition an identity field.  In general, for a
superposition of fundamental boundary states, the associated open string
theory has more than one identity fields.}  Only those closed
string states that appear in the boundary state have a non-zero
coupling.  Ishibashi states that are present in the fundamental
boundary states may be canceled out in the summation process as for
instance in the $N$ $D0$-branes configuration.  This
misleadingly suggests that certain closed string fields do not couple to
open string identity fields and thus leads to the violation of the cluster
condition, one of the sewing constraints \lew\ required for
a local boundary conformal field theory (BCFT).  With the use of the CP
index, all couplings of closed string fields to the open string identity
fields are explicitly present in the boundary state and the
cluster condition can be verified.  This subtlety
can be overlooked if the boundary state is used solely to calculate the
open string spectrum and test Cardy's condition.  Thus, we see that the
orbifold mechanism functions at the level of summation and does not
distinguish whether  the orbifold boundary state is a fundamental or a
superposition of fundamental states.

\medskip
\noindent{\it c.~ $r=M/N$ case}

The geometry of the orbifold $D0$-brane boundary state in this case is made
trivial by following the descriptions of the two preceding cases. 
One can first orbifold by
$r=\frac{1}{N}$ followed by another orbifold by $r=M$.  The net result
on the $R=M/N$ orbifold space is $M$ equally-spaced $D0$-branes with a
separation of $\Delta\xo=\frac{2\pi}{N}$.  Notice that on the orbifold
space, the original shift symmetry in the $R=1$ theory,
$\xo\to\xo+2\pi$, is still
present - that is there is a $D0$-brane located a distance $\pm2\pi$ apart
from any single $D0$-brane.  Indeed, the original symmetry of the theory
is preserved under the orbifold.\foot{This becomes manifest by working in the
{\bf R}$^1$ covering space of $S^1$.  A $D0$-brane situated at
$\xo\!=\!0$ in the $R\!=\!1$ theory corresponds to a $D0$-brane located
at each point $2\pi\bZ$ on {\bf R}$^1$.  Applying the orbifold action,
$D0$-branes in {\bf R}$^1$ are now located at $2\pi(m+rn)$ for
$m,n\in\bZ$.  This configuration corresponds in the orbifold space to
$D0$-branes located at ${2\pi(m+rn)~{\rm mod}~2\pi r}={2\pi m~{\rm mod}~
2\pi r}$, which preserves the original $2\pi$ shift symmetry.}  This implies the presence of
conformal dimension one winding states that stretch a distance of
$2\pi$.  In addition to the translation field $i\partial
X$, the marginal winding states together provide two additional truly
marginal boundary fields that can be used to
deform the boundary state.\foot{It is worth noting that a conformal
dimension one winding field that stretches a distance $2\pi$ by itself
is not truly marginal (see \rs\ for conditions on true marginality).  In
fact, it is precisely because the $D0$-branes are equally-spaced that
we can construct two truly marginal winding fields.  The two
are linear combinations of all dimension one winding fields.}  A
deformation by the truly marginal winding fields will lead to the
orbifold Virasoro boundary state which from \overl\ and \overla\ can
be seen to have three open strings fields of dimension one.  The
presence of truly marginal fields allows all bulk orbifold boundary states to
be connected by marginal deformations.
  
\medskip
\noindent{\it d.~ $r=irrational$ case}

The group action $\Gamma$ for irrational $r$ is of
infinite order and the orbifold boundary states given in \pbs\ is 
exactly those given in \refs{\gr, \jan}.  It is interesting to point out
that the orbifold boundary state only consists of closed string states
from the conformal family of the identity field which are present in the
closed string theory at any radius.  As for the geometry of the
orbifold $D0$-brane boundary state, it must preserve the symmetry of the
original shift symmetry.  This implies that the orbifold boundary state
at $R=r$ is
geometrically that of $D0$-branes located at points $p={2\pi m~{\rm
mod}~ 2\pi r}$.  This
gives a dense set of $D0$-branes and explains why the boundary
entropy, that is after ``summing'' over the contributions from the infinite
number of $D0$-branes present, is infinite.  More significantly, the
Virasoro boundary states, obtained by deforming with
truly marginal boundary fields and suspected to be fundamental, must
also have infinite boundary entropy.  Since the mass of the brane is
proportional to the boundary entropy \hkms , we find that the irrational radius
Virasoro boundary states have infinite mass, and therefore, are not
likely to be physically relevant.\foot{One may wonder what happens to
the finite mass Virasoro boundary states at
rational radius when perturbed by the marginal closed
string vertex operator $\partial X {\bar\partial}X$ to irrational
radius.  Indeed, the Virasoro boundary condition under this
perturbation becomes non-conformal.  For $R=\frac{1}{2}$, Sen has
explicitly shown in \sen\ that a non-zero tadpole arises at first order
in perturbation.}

\newsec{Boundary State Orbifold as Open String Quotient}
 
We have applied an asymmetric orbifold on the boundary states at $R=1$ to
obtain boundary states at $R=r$.  This is a closed string description
mapping boundary conditions from an open string theory at a particular
radius to those at a different radius by means of an orbifold.  As might be
expected, there is an open string description too.  In the open string
sector, different boundary conditions are distinguished by their
associated vertex operator algebras.  The boundary orbifold mechanism
suggests that it is possible to generate the vertex operator algebras at
$R=r$ from those at $R=1$.  Indeed, this can be accomplished using
the quotient algebra construction \refs{\dou, \mamo} which
has been used successfully to describe open strings on symmetric
orbifolds \refs{\gp, \dm}.  

To describe the vertex operator algebra at $R=r$, we
need to study how the group action $\Gamma$ acts on the vertex
operators $V(X, \tX)$ at $R=1$.  Of course, $\Gamma$ must act on the
fields, $X$ and $\tX$, as given in \geh .  But notice that in \pbs , $\Gamma$
maps out the images of the orbifold boundary state in the covering space.
With a superposition of boundary states in the covering space, the open
string sector covering space algebra $V(X, \tX)$ must be
matrix-valued and labelled by two CP indices.  Thus, $\Gamma$ will act
on the CP index with an action $\gamma(h)$ dependent on the representation
space chosen for the CP indices.  For obtaining the bulk boundary states,
such as the Virasoro orbifold boundary states of \pbs , the regular
representation
is required with CP indices $i,j=0,\ldots,{|\Gamma|-1}$.  A valid
representation with a smaller dimension will give the vertex algebra for
a fractional boundary state.  The quotient algebra or the
algebra of vertex operators at $R=r$ is then the matrix-valued algebra
$V(X, \tX)$ satisfying the equivalence relation condition 
\eqn\covrep{\gamma(h)\,V(X,\tX)\,\gamma^{-1}(h) = V(h(X,\tX))
\quad\quad {\rm for~all}~~h\in \Gamma~.}
The quotient algebra is in fact universal in that it may be applied to
any group action $\Gamma$ with an unitary representation $\gamma (h)$
acting on a Hilbert space ${\cal H}$.  

\subsec{Example: $r=2$}

We will work out the vertex
algebras for $D1$- and $D0$-brane boundary conditions for $r=2$ and will not
consider non-U(1) boundary conditions.  The quotient algebra
reproduces the $R=2$ vertex algebra of a single $D1$-brane for
the $D1$-brane boundary condition and two equally-spaced $D0$-branes for the
$D0$-brane boundary condition as required from the results of the
boundary state orbifold \pbs.  For simplicity, we will focus the
analysis on the open string tachyon vertex operators $T(X,\tX)$.  Since
the oscillator modes
are not affected by the orbifold, our analysis can be easily extended to
include all vertex operators.  Before proceeding, we note that open
string vertex operators are situated at the boundary $z\!=\!\bz\,$, and
in particular, those of the tachyons have the form
$e^{i\frac{n}{R}X(z,\bz)}$ and $e^{imR\tX(z,\bz)}$, respectively, for a
$D1-$ and a $D0$-brane at radius $R$.  We emphasize that $X$ and $\tX$
have different mode expansions for different boundary conditions.  The
operator product algebra can be obtained by applying Wick contractions
with brane specific Green's functions:
${<X(z_1,\bz_1)X(z_2,\bz_2)>_{D1}}={<\tX(z_1,\bz_1)\tX(z_2,\bz_2)>_{D0}}=-\frac{\alpha'}{2}\ln|z_1-z_2|^2-\frac{\alpha'}{2}\ln|z_1-\bz_2|^2$.

We first analyze the $D1$-brane case.  For the quotient algebra, we have
$|\Gamma|=2$ and the nontrivial action on the CP index is given by
$\gamma(h_{0,1})=\left (\matrix{ 0 & 1\cr 1 & 0} \right )\equiv\gamma$.
Here, the regular representation is the minimal
representation because the action $h_{0,1}$ (defined in \geh ) on
the covering space maps a $D1$-brane to another $D1$-brane with a
distinct value of the constant gauge potential.  The tachyon vertex operator at
$R=1$ is a $2\times 2$ matrix with vertex operators of the form $e^{inX}$ for
diagonal elements and $e^{i(n+\frac{1}{2})X}$ for off-diagonal elements.
The half-integer momenta are required for the off-diagonal elements since these
are open strings stretching between two $D1$-branes ``separated'' by
$\Delta\txo=\pi$.  Notice also that $\Gamma$ does not act on the field
$X$ in the vertex operator.  All together, the tachyon vertex operators that
satisfies \covrep\ are expressed as
\eqn\tneu{T(X)=\sum_{n} a_{n}\, e^{i n X}\left( \matrix{1&0\cr
0&1}\right ) + \sum_{n} b_{n}\, e^{i(n+\frac{1}{2}) X} \left (
\matrix{0&1\cr 1&0} \right )~,} 
where all $a$'s and $b$'s denote constant coefficients.  Indeed, \tneu\
is algebraically equivalent to the expression for the tachyon vertex
operator of a single $D1$-brane at $R=2$, $\sum_n a'_{n}\,e^{i\frac{n}{2}X}$. 

We now consider the $D0$-brane case.  The nontrivial group
action $h_{0,1}$ only acts on the vertex operator by shifting $\tX$ and
does not change the location of the $D0$-brane.  Keeping the regular
representation with $\gamma(h_{0,1})=\gamma$, we have on the 
covering space two coincident $D0$-branes.  Thus, the 
tachyon vertex operator at $R=1$ have the form $e^{im\tX}$ for
both diagonal and off-diagonal elements.  The equivalence relation
\covrep\ requires the general expression for the operator to be 
\eqn\tdip{\eqalign{T(\tX)=&{\sum_{m} a_{m} \left
(\matrix{e^{im\tX}&0\cr 0& e^{im(\tX+\pi)}} \right )} + { \sum_{m}
b_{m} \left (\matrix{0&e^{im\tX} \cr e^{im(\tX+\pi)}& 0 } \right
)}\cr=&\sum_{m}\sum_{k,l=0,1} a_{m\,k\,l} \;e^{i(2m+k)\tX}{\left (\matrix{1&0\cr
0&-1} \right )}^{k}{\left (\matrix{0&1\cr 1&0}\right )}^l ~, }} 
where a matrix raised to the zeroth power is
taken to be the identity matrix.  In the second line we have expressed
the tachyon operator in the natural matrix basis.  This quotient algebra
should be compared to that of two equally spaced $D0$-branes at $R=2$
which can be expressed as 
\eqn\tdia{\eqalign{T'(\tX)=&\left (\matrix{\sum_{m} a'_{m}\,
e^{i2m\tX}&\sum_{m}b'_{m}\, e^{i(2m+1)\tX}\cr \sum_{m}c'_{m}\,
e^{i(2m+1)\tX}&\sum_{m} d'_{m}\, e^{i2m\tX}}\right )\cr=&\sum_{m}\sum_{k,l=0,1} a_{m\,k\,l}' \;e^{i(2m+l)\tX}{\left (\matrix{1&0\cr
0&-1} \right )}^{k}{\left (\matrix{0&1\cr 1&0}\right )}^l ~, }} 
where in the second line, the tachyon operator is expressed in the
natural basis of the quotient algebra.  Comparing
\tdip\ with \tdia , we see that both have the same open
string spectrum and indeed are equivalent representations of the same algebra. 
An explicit relationship that connects the two representations is given
by the identification $a_{m00}=a'_{m00}\,$, $a_{m01}=a'_{m10}\,$, 
$a_{m10}=a'_{m01}\,$, and $a_{m11}=-a'_{m11}\,$.  Thus, the quotient algebra
and the vertex algebra on the orbifold space are identical up to a
change of basis.

For the $D0$-brane boundary condition, it is also possible to
express $\gamma(h_{0,1})$ in the one-dimensional irreducible
representation.  The irreducible
representation can be used here because the location of the $D0$-brane
is invariant under $\Gamma$ in the $R=1$ covering space.  The equivalence
condition \covrep\ for a single irreducible representation becomes
$T(\tX)=T(\tX+\pi)$.  The quotient algebra thus have elements
$T(\tX)=\sum_m a_m\,e^{i2m\tX}$, which are exactly the elements of the tachyon vertex algebra for a single $D0$-brane in the $R=2$ theory.  Thus, we see
that for $r=2$, the irreducible representation corresponds to the
fractional brane in the orbifold theory.

\subsec{Generic Radius and the Noncommutative Weyl Algebra} 

For a generic $r$, the quotient algebra constructed using the regular
representation with dimension $|\Gamma|={(m_{max}'+1)(n_{max}'+1)}$ will
produce the expected open string vertex operator algebra for the
orbifold boundary state in \pbs .  For the $D1$- and $D0$-brane
boundary conditions, the generalization of the above $r=2$ example is
straightforward.  Below, we will describe some general features of the
quotient algebra applicable for these two boundary conditions.

For $D1$- and $D0$-brane boundary conditions, the vertex operators have
dependence only on $X$ and $\tX$, respectively.  Moreover, with the
group action being a tensor product,
$\Gamma={\bZ}_{(m'_{max}+1)}\times{\bZ}_{(n'_{max}+1)}\,$, it is natural
to express the regular representation labelled by $j$ as a direct
product representation labelled by two indices $j=(j_1,j_2)$ with
$j_1=0,\ldots,m'_{max}$ and $j_2=0,\dots,n'_{max}$.  The regular
representation corresponds to
$|\Gamma|$ number of branes on the covering space.  These branes are
labelled by $(j_1,j_2)$ and the role they play in constructing the quotient
algebra can be gleaned from the actions of ${\bZ}_{(m'_{max}+1)}$ and
${\bZ}_{(n'_{max}+1)}$ and from insights obtained from the $r=2$ example.  

Let us consider the $D0$-brane boundary condition.  The
${\bZ}_{(m'_{max}+1)}$ action (acting on $X$) generates $m'_{max}+1$
equally-spaced $D0$-branes on the
covering space. Thus, $j_1$ labels $D0$-branes that are separated on the
covering space.  Similar to the $D1$-brane boundary condition for $r=2$, the
separated branes provide for the open string spectrum of a single
$D0$-brane on the orbifold space.  In contrast, the
${\bZ}_{(n'_{max}+1)}$ action (acting on $\tX$) does not affect
the location of the $D0$ branes; therefore, $j_2$ labels the
$n'_{max}+1$ coincident $D0$-branes situated at each image point.
Like the $D0$-brane boundary condition for $r=2$,
$n'_{max}+1$ coincident $D0$-branes on each image point on the covering
space will give, after applying the equivalence relation, the vertex
algebra of $n'_{max}+1$ equally-spaced $D0$-branes on the orbifold
space.  Having only one $D0$-brane on each image point on the covering space
corresponds to only one $D0$-brane on the orbifold space.  Thus, we note
that the fractional brane for the $D0$-brane boundary condition is
associated with the representation with dimension $m_{max}'+1$.  The
analysis for the $D1$-brane boundary condition is identical to that of the
$D0$-brane except for exchanging $m'_{max}\leftrightarrow n'_{max}$ and
$j_1\leftrightarrow j_2$.

The quotient algebra for $D1$- and $D0$-brane boundary conditions in the direct
product representation $(j_1,j_2)$ is manifestly a direct product
algebra.  For ease of discussion, we will focus on the tachyon vertex
operator algebra and treat it as a matrix algebra by ignoring the
contribution from the Green's function.  Again, we first consider the
$D0$-brane boundary condition.  The subalgebra associated with the
$j_1$ index is commutative and similar to that given in \tneu .  It is
generated by the shift matrix $V|j_1\!>=|j_1-1\!>$ with $V$ raised to
the $(m'_{max}+1)$-th power being the identity matrix.  The subalgebra
associated with the $j_2$ index like that of \tdip\ is noncommutative
and is generated by both the shift matrix $V$ and the clock matrix,
$U|j_2\!>=w^{j_2}|j_2\!>$ with $w=e^{i\frac{2\pi}{r}}$. 
Indeed, the subalgebra labelled by $j_2$ is the Weyl algebra,
$UV=w^{-1}VU\,$.\foot{The presence of a noncommutative algebra at
irrational radius $R$ was hinted at by Friedan \fri .  Also, the
appearance of the Weyl algebra in quotient algebras has been much
exploited in the field of noncommutative geometry \conn .  See, for
example, \dou\ for a concise exposition from a physical perspective.}
As an explicit example, the tachyon vertex operator satisfying the
equivalence relation \covrep\ for $r=M/N$ can be expressed as 
\eqn\tdiea{
T(\tX)=\sum_{m}\sum_{n=0}^{m'_{max}}\sum_{k,l=0}^{n'_{max}} a_{m\,n\,k\,l}
\;e^{i\,nr\tX}V'^n \otimes \; e^{i(Mm+k)\tX}U^k\ V^l~, } 
where $V'$ denotes the shift operator acting on the $j_1$ index and $U$
and $V$ act on the $j_2$ index.  With $j_2=0,\ldots,n'_{max}$, the Weyl
subalgebra is present only if
$n'_{max}\ge 1$.  Hence, on the orbifold space, there must be two or more
equally-spaced $D0$-branes.  For the $D1$-brane boundary condition, the
Weyl subalgebra is associated with the $j_1$ index with $w=e^{i 2\pi
r}$.  In general, the Weyl subalgebra is present only if the quotient
algebra is associated with an orbifold boundary state that is not
fundamental (i.e. not a Virasoro brane nor a single $D0$- or $D1$-brane). 

The appearance of the Weyl algebra can be understood
from the orbifold space perspective.  As we have argued, the quotient
algebra is identical to the vertex algebra associated with the
orbifold boundary state.  This implies that the vertex algebra for two or more
equally-spaced $D0$- or $D1$-branes contains the Weyl algebra. Indeed,
we have seen this in the $r=2$ example, where the vertex operators of two
equally-spaced $D0$-branes can be expressed in the clock and shift basis
as in \tdia .  With equally-spaced branes, each element in the matrix
vertex algebra and those diagonally above and below it have the same set of
allowed momentum.  Because of this, one can always express the vertex
operator matrix using the complete matrix basis generated by $U$ and $V$.
In this basis, the Weyl algebra becomes manifest. 

\newsec{Discussion}

The asymmetric shift orbifold that we have presented in fact connects two
arbitrary points on the moduli space of the closed string theory on a circle.
We have chosen the covering theory to be situated at $R=1$ only to
facilitate the construction of the Virasoro boundary states at other radii.
In doing so, all Virasoro boundary states from the orbifold are bulk
boundary states and can be easily constructed without considering
twisted sector closed string states.  As noted, in the open string description,
bulk boundary states correspond to the regular representation in the
quotient algebra construction.
\smallskip
Bulk orbifold boundary states are interesting objects.  Neglecting the
normalization factor $\sqrt{|\Gamma|}$, the boundary state is present in both
the covering and the orbifold theory.  The bulk orbifold boundary state in the
{\it covering theory} is a superposition of boundary states; in the
orbifold theory, we have seen that it may be either a fundamental
boundary state or also a superposition of boundary states.  That it can
be found within two different closed string theories is the direct result that
the spin zero closed string states that make up the boundary state are
present in both closed string theories.  Indeed, the orbifold action
has provided us a systematic mechanism to identify the set of
shared closed string states between the two theories.  Since each
boundary state corresponds to a BCFT, a bulk orbifold boundary state
effectively relates two BCFTs situated within two different bulk theories.
\smallskip
We speculate that a generalization of the bulk orbifold boundary state
may be made even when two closed string theories are not connected by an
orbifold but share a set of spin zero closed strings.  We point
out that boundary states are typically constructed using only a
subset of the set of spin zero closed string states.  Thus, in general,
consider two closed
string theories, labelled by $A$ and $A'$, that have in common a set of
states called $S$.  Now if a boundary state $|B\dr$ that satisfy Cardy's
condition (at least with itself) can 
be constructed from states in $S$ in theory $A$, then $|B\dr$ up to a
normalization is also a boundary state in theory $A'$.  The presence of
$|B\dr$ effectively gives a
relationship between two boundary field theories associated with different
bulk field theories.  Specifically, the boundary field spectrum of the
two BCFTs, possibly up to an integer multiplicity factor, are identical
by Cardy's condition.  This generalization raises the possibility that 
two BCFTs may be related although there may not be a relation between
their associated bulk conformal field theories.
\smallskip
We should mention that throughout the paper, we have neglected the
sewing constraints that must be satisfied for the local consistency of
any BCFT.  We will only add that
one of the sewing constraints, the cluster condition, may be helpful in
determining whether an orbifold boundary state is fundamental.  As noted
earlier, the construction of orbifold boundary states is at the level of
``summation'' and does not provide any information on whether the
orbifold boundary state is fundamental or a superposition of
fundamental boundary states.  However, if the boundary state does not
satisfy the cluster relation, then it can not be fundamental.

\bigskip\medskip\noindent 
{\bf Acknowledgements:}
I am grateful to J. Harvey for helpful discussions and critical
comments on the manuscript.  I would also like to thank R. Bao,
B. Carneiro da Cunha, B. Craps, D. Kutasov, E. Martinec, and D. Sahakyan
for useful discussions.  I am also grateful to the referee for 
valuable comments on the manuscript.  This work was supported in part by
NSF grant PHY-9901194. 

\listrefs
\end